\documentclass[sn-mathphys,Numbered]{sn-jnl}

\usepackage{graphicx}%
\usepackage{multirow}%
\usepackage{amsmath,amssymb,amsfonts}%
\usepackage{amsthm}%
\usepackage{mathrsfs}%
\usepackage[title]{appendix}%
\usepackage{xcolor}%
\usepackage{textcomp}%
\usepackage{manyfoot}%
\usepackage{booktabs}%
\usepackage{algorithm}%
\usepackage{algorithmicx}%
\usepackage{algpseudocode}%
\usepackage{listings}%

\theoremstyle{thmstyleone}%
\theoremstyle{thmstyletwo}%

\theoremstyle{thmstylethree}%

\begin{document}

\title[Article Title]{Quantum-inspired anomaly detection, a QUBO formulation}


\author*[1]{\fnm{Julien} \sur{Mellaerts}}\email{julien.mellaerts@eviden.com}

\affil*[1]{\orgname{Eviden}, \orgaddress{\city{Les Clayes-sous-Bois}, \country{France}}}


\abstract{Anomaly detection is a crucial task in machine learning that involves identifying unusual patterns or events in data. It has numerous applications in various domains such as finance, healthcare, and cybersecurity. With the advent of quantum computing, there has been a growing interest in developing quantum approaches to anomaly detection. After reviewing traditional approaches to anomaly detection relying on statistical or distance-based methods, we will propose a Quadratic Unconstrained Binary Optimization (QUBO) model formulation of anomaly detection, compare it with classical methods, and discuss its scalability on current Quantum Processing Units (QPU).}




\maketitle

\section{Introduction}\label{sec1}

Quantum computing and machine learning are two rapidly advancing fields that have the potential to revolutionize various domains. Quantum machine learning (QML) combines the power of quantum computing with the principles of machine learning, opening up new possibilities for solving complex problems and enhancing the capabilities of traditional machine learning algorithms. At its core, quantum computing harnesses the principles of quantum mechanics to manipulate and process information in ways that classical computers cannot. Quantum bits, or qubits, can exist in a superposition of states, allowing for parallel computation and the potential to perform complex calculations exponentially faster than classical computers for certain problems. This unique property of quantum computing forms the basis for developing quantum machine learning algorithms.

\bigskip\noindent
Quantum machine learning aims to leverage quantum computers to enhance the speed and efficiency of various machine learning tasks. It offers the promise of solving computationally intensive problems that are intractable for classical machine learning algorithms, and relies on phase estimation, quantum matrix inversion and amplitude amplification methods \citep{bib1, bib2, bib3, bib4}. Several key areas within quantum machine learning have already shown promising advancements, notably in data classification, graphs applications or pattern recognition \citep{bib5, bib6}.

\bigskip\noindent
Anomaly detection refers to the identification of patterns or instances that deviate significantly from the expected or normal behavior within a data set. These anomalies often indicate potential fraud, errors, faults, or other exceptional events that require attention. By leveraging advanced algorithms and statistical techniques, anomaly detection algorithms are widely used in many applications like finance, healthcare and cybersecurity \citep{bib7, bib8}.

\bigskip\noindent
Current Quantum Anomaly Detection (QAD) algorithms are implemented on gate-based quantum computing paradigm making use of amplitude estimation, variational or quantum K-means technics \citep{bib9, bib10, bib11, bib12}. Current Noisy Intermediate-Scale Quantum (NISQ) computers are composed of number-limited noisy qubits, impacting the practicability of these gate-based QAD algorithms in terms of data set size or dimensionality.

\bigskip\noindent
In this paper, I propose a Quadratic Unconstrained Binary Optimization (QUBO) formulation of anomaly detection, benchmark it against classical methods on real-world data sets, and finally discuss on its scalability and effectivness.

\section{Classical anomaly detection algorithms}\label{sec2}

There are several types of anomaly detection algorithms, including statistical-based methods, density-based methods, and machine learning-based methods. Statistical-based anomaly detection algorithms assume that normal data points follow a known statistical distribution, such as Gaussian (normal) distribution. These methods use statistical techniques to model the data distribution and identify instances that fall outside the expected range. Common statistical-based algorithms include:
\begin{unenumerate}[1.]
\item  Z-Score: This algorithm measures the distance of each data point from the mean in terms of standard deviations. Points that fall above or below a certain threshold are considered anomalies.

\item Gaussian Mixture Models (GMM): GMM assumes that the data is a combination of several Gaussian distributions. It estimates the parameters of the Gaussian components and identifies data points with low probability as anomalies.

\item Box Plot: A box plot summarizes the distribution of a data set by representing quartiles, outliers (anomalies), and extreme values. Data points beyond a certain threshold are considered outliers or anomalies.
\end{unenumerate}

Density-based anomaly detection algorithms focus on the proximity or density of data points in the feature space. They assume that anomalies are far from their neighboring points or are in low-density regions. Common density-based algorithms include:
\begin{unenumerate}[2.]
\item  k-Nearest Neighbors (k-NN): k-NN identifies anomalies based on the distance to their k-nearest neighbors. Points that have few nearby neighbors or have significantly larger distances are considered anomalies.

\item Local Outlier Factor (LOF): LOF measures the local density of data points compared to their neighbors. Anomalies have lower local densities compared to their neighbors.

\item Density-Based Spatial Clustering of Applications with Noise (DBSCAN): DBSCAN groups data points based on their density and identifies outliers as points that do not belong to any cluster.
\end{unenumerate}

Machine learning-based anomaly detection algorithms leverage advanced techniques to learn the underlying patterns and characteristics of normal data. They aim to build models that can differentiate between normal and anomalous instances. Common machine learning-based algorithms include:
\begin{unenumerate}[3.]
\item  Support Vector Machines (SVM): SVM separates the data into different classes and identifies instances that fall on the "wrong" side of the separation boundary as anomalies.

\item Isolation Forest: This algorithm constructs an ensemble of isolation trees to isolate anomalies efficiently. Anomalies require fewer splits to be isolated compared to normal instances.

\item Neural Networks: Neural networks can learn complex patterns in the data and detect anomalies based on deviations from the learned representation.

\item Autoencoders: Autoencoders are neural network architectures that aim to reconstruct input data. Anomalies lead to higher reconstruction errors, allowing for their detection.
\end{unenumerate}

The idea behind the proposed QUBO formulation of anomaly detection algorithm is to combine both statistical and density-based methods in order to increase the accuracy of the algorithm.

\section{Quantum-inspired anomaly detection}\label{sec3}

In this section, I describe the proposed quantum-inspired anomaly detection algorithm using a QUBO formulation. I start with a general definition of QUBO and translate it for anomaly detection.

\subsection{QUBO formulation}\label{subsec2}

Quadratic Unconstrained Binary Optimization (QUBO) is a mathematical optimization problem that involves minimizing a quadratic objective function subject to binary variables. In other words, it is an optimization problem where the goal is to find the binary values of a set of variables that minimize a quadratic cost function.

The binary variables in QUBO can take only two values: 0 and 1, representing a binary choice or decision. The objective function in QUBO is quadratic, meaning it consists of quadratic terms involving the binary variables and possibly linear terms as well. The goal is to find the assignment of binary values to the variables that minimizes the overall value of the objective function.

A QUBO model is expressed as an optimization problem:
\begin{equation}
f(x) = \sum\limits_{i=0}^{N-1}Q_{i,i}x_i + \sum\limits_{i,j=0, i \neq j}^{N-1}Q_{i,j}x_ix_j
\end{equation}

Diagonal terms of Q represent linear terms whereas off-diagonal terms represent quadratic terms. 

\bigskip
A parallel can be drawn between QUBO formulation and anomaly detection algorithms, in the sense that, statistical-based algorithms can be seen as finding most distant data points from the centroid of a Gaussian distributed data set, which is relative to linear terms of the QUBO, and density-based algorithms can be seen as finding neighbors most distant data points, which is relative to quadratic terms of the QUBO. This drawing allows the combination of both statistical and density based algorithms.

\bigskip
We can now give a more rigorous formulation of the problem. Given a data set $X = (x_0, ..., x_{N-1})$ of $N$ data points, let's define $d_{i}$ the distance between a point $x_i \in X$ and the centroid of the data set distribution, and $d_{i,j}$ the distance between two data points $x_i, x_j \in X$ with $i \neq j$, the cost function to maximize is: 

\begin{equation}
Q(x,\alpha) = \alpha \sum\limits_{i=0}^{N-1}d_ix_i + (1-\alpha)\sum\limits_{i,j=0, i \neq j}^{N-1}d_{i,j}x_ix_j
\end{equation}
subject to:

$\sum\limits_{i=0}^{N-1}x_i=k$

\bigskip\noindent
where:

$0 \leq \alpha \leq 1$, is a weighting parameter for linear and quadratic terms.

$0 < k \leq N$ is the number of outliers in the data set.

\bigskip
The $\alpha$ parameter is specific for each data set and needs to be learnt during a training phase. 

\bigskip
The number of outliers to select in the data set can be constrained in the QUBO by adding a penalty to the cost function:
\begin{equation}
-A(\sum\limits_{i=0}^{N-1}x_i-k)^2
\end{equation}
where:

$A$ is a penalty weight that must be scaled accordingly to QUBO terms so that $A>max(Q)$.

\bigskip
To get most accurate results, it is important to limit the number of quadratic terms of each variable to the number of outliers, by filling only the k-furthest neighbors.

The chosen distance metric is the Mahalanobis distance, which is an effective multivariate distance metric available in many machine learning models, including those that will be used to benchmark anomaly detection methods.

\subsection{Benchmarks}\label{subsec3}
In this section I present benchmarking methods and results. For all benchmarks, I compare quantum-inspired anomaly detection to :
\begin{unenumerate}[4.]
\item  Robust covariance: sklearn.covariance.EllipticEnvelope

\item One-Class SVM: sklearn.vm.OneClassSVM

\item One-Class SVM (SGD): sklearn.linearmodel.SGDOneClassSVM

\item Isolation Forest: sklearn.ensemble.IsolationForest

\item Local Outlier Factor: sklearn.neighbors.LocalOutlierFactor
\end{unenumerate}

Each benchmark instance is solved using qbsolv, and the metric is the Area Under the Receiver Operating Characteristic Curve (ROC AUC) score. 

\subsubsection{Random Gaussian distributed data points}\label{subsubsec1}

In this first benchmark, the quantum-inspired anomaly detection is compared to classical methods on three different random Gaussian distributed data points, with three different standard deviations.

\begin{figure}[h]%
\centering
\includegraphics[width=0.9\textwidth]{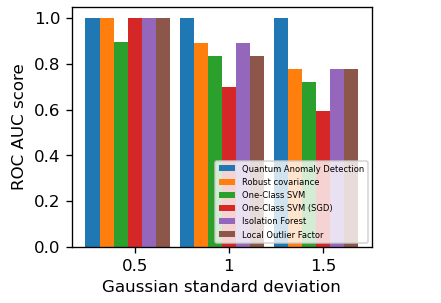}
\caption{Random Gaussian distributed data points ROC AUC scores of anomaly detection algorithms for three different standard deviations.}\label{fig1}
\end{figure}

\subsubsection{MNIST data set}\label{subsubsec2}
The MNIST data set contains 60,000 training images and 10,000 test images
of handwritten digits and is widely used in classification benchmarks.

In this second benchmark, I compare quantum-inspired anomaly detection to classical methods on three different MNIST data set configurations:
\begin{unenumerate}[5.]
\item  45 samples of handwritten 0, 5 samples of handwritten 9

\item 45 samples of handwritten 7, 5 samples of handwritten 1

\item 45 samples of handwritten 2, 5 samples of handwritten 3
\end{unenumerate}

Minority of handwritten digit of each configuration should be identified as outlier by anomaly detection algorithms.

\begin{figure}[h]%
\centering
\includegraphics[width=0.9\textwidth]{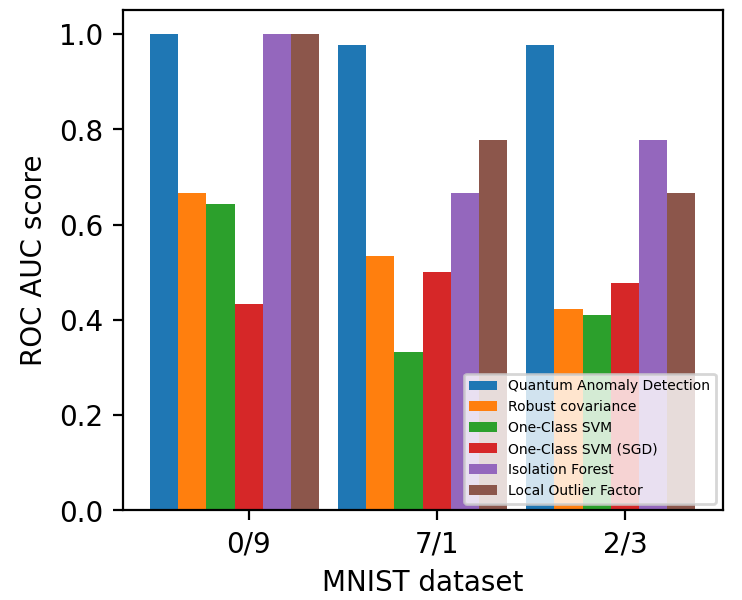}
\caption{MNIST data set ROC AUC scores of anomaly detection algorithms for three different configurations.}\label{fig2}
\end{figure}

\subsubsection{Credit card fraud detection}\label{subsubsec3}
In this last benchmark, I compare quantum-inspired anomaly detection to classical methods on credit card fraud detection data set \cite{bib13}.

The data set contains credit card transactions that occured in two days in September 2013 by european cardholders, there are 492 frauds out of 284,807 transactions. The data set is highly unbalanced, the positive class (frauds) account for 0.172 percent of all transactions.

\begin{figure}[h]%
\centering
\includegraphics[width=0.9\textwidth]{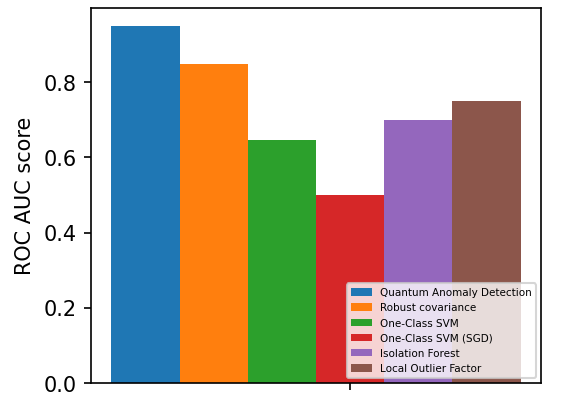}
\caption{Credit card fraud data set ROC AUC scores of anomaly detection algorithms.}\label{fig3}
\end{figure}

\section{Conclusion}\label{sec4}

In this article, I present a novel QUBO formulation for anomaly detection which shows improved accuracy in comparison with classical methods on random data sets as well as real-word data sets. Limiting quadratic terms to k-furthest neighbors improves accuracy and potentially enables finding an embedding on actual QPUs, however the outlier selection penalty (equation 3) imposes all-to-all quadratic interactions, making the QUBO not solvable by near-term QPUs. An other method of restricting outlier selection should be used to allow this problem to be solved on connectivity-limited QPUs. Nevertheless, this quantum-inspired anomaly detection algorithm can be solved without hardware constraint and with high number of variables using simulated quantum annealing, simulated annealing, or any QUBO solver.

\backmatter





\bibliography{sn-bibliography}

\end{document}